\documentclass[12pt, a4paper]{article}
\pdfoutput=1

\usepackage{multirow}
\usepackage{amsmath}
\usepackage{amsfonts}
\usepackage{amssymb}
\usepackage{graphicx, rotating}
\usepackage{epstopdf}
\usepackage{epsfig}
\usepackage{latexsym}
\usepackage{graphicx}
\usepackage{color}
\usepackage{amsmath,amssymb}
\usepackage{cite}
\usepackage{slashed}

\ifx\pdfoutput\undefined
\usepackage[dvips,bookmarks=false]{hyperref}	
\else
\usepackage{hyperref}	
\fi

\hypersetup{colorlinks, citecolor=bluscuro, linkcolor=black, urlcolor=bluscuro}
\definecolor{rossos}{cmyk}{0,1,1,0.55}
\definecolor{bluscuro}{rgb}{0.15, 0.2, .85}
\definecolor{bluchiaro}{cmyk}{1,.3,0.,0.1}

\newcommand{\eq}[1]{Eq.~(\ref{#1})}

\def\lra#1{\overset{\text{\scriptsize$\leftrightarrow$}}{#1}}

\newcommand{\beq}{\begin{equation}} 
\newcommand{\eeq}{\end{equation}} 
\newcommand{\be}{\begin{equation}}
\newcommand{\ee}{\end{equation}}
\newcommand{\ba}{\begin{array}}  
\newcommand{\ea}{\end{array}} 
\newcommand{\bea}{\begin{eqnarray}}  
\newcommand{\eea}{\end{eqnarray} }  
\newcommand{\bal}{\begin{align}}
\newcommand{\eal}{\end{align}}   
\newcommand{\bi}{\begin{itemize}}  
\newcommand{\ei}{\end{itemize}}  
\newcommand{\ben}{\begin{enumerate}}  
\newcommand{\een}{\end{enumerate}}  
\newcommand{\bc}{\begin{center}}
\newcommand{\ec}{\end{center}} 
\newcommand{\bt}{\begin{table}}
\newcommand{\et}{\end{table}}  
\newcommand{\btb}{\begin{tabular}}
\newcommand{\etb}{\end{tabular}}

\begin{document}

\begin{titlepage}

\vspace*{-2cm}
\begin{flushright}
CERN-TH-2016-082\\
DESY 16-067\\
LPT Orsay 16-32 \\
\vspace*{2mm}
\end{flushright}

\begin{center}
\vspace*{15mm}

\vspace{1cm}
{\LARGE \bf
On the Validity of  the Effective Field Theory \\[0.25cm] Approach  to SM Precision Tests
} \\
\vspace{1.4cm}

\renewcommand{\thefootnote}{\fnsymbol{footnote}}
{\bf Roberto Contino$^{a,b}$\footnote{\hspace{0.1cm}On leave of absence from Universit\`a di Roma La Sapienza and INFN, Roma, Italy.}, Adam Falkowski$^c$, Florian Goertz$^b$,\\ Christophe Grojean$^{d}$\footnote{\hspace{0.1cm}On  leave  of absence from  ICREA,  E-08010  Barcelona,  Spain.}, Francesco Riva$^b$}
\renewcommand{\thefootnote}{\arabic{footnote}}
\setcounter{footnote}{0}

\vspace*{.5cm}
\centerline{ $^a${\it Institut de Th\'eorie des Ph\'enomenes Physiques, EPFL, Lausanne, Switzerland}}
\centerline{$^b${\it Theoretical Physics Department, CERN, Geneva, Switzerland}}
\centerline{$^c${\it Laboratoire de Physique Th\'{e}orique, Bat.~210, Universit\'{e} Paris-Sud, 91405 Orsay, France}}
\centerline{$^d${\it DESY,  Notkestrasse  85,  D-22607  Hamburg,  Germany}}

\vspace*{.2cm}

\end{center}

\vspace*{10mm}
\begin{abstract}\noindent\normalsize
We discuss the conditions for an effective field theory (EFT) to give an adequate low-energy description of an underlying physics beyond the Standard Model (SM).
Starting from the EFT where the SM is extended by dimension-6 operators, experimental data can be used without further assumptions to 
measure (or set limits on) the EFT parameters.
The interpretation of these results requires instead a set of broad assumptions ({\it e.g.} power counting rules) on the UV dynamics.
This allows one to establish, in a bottom-up approach, the validity range of the EFT description, and to assess the error 
associated with the truncation of the EFT series.
We give a practical prescription on how experimental results could be reported, so that they admit a 
maximally broad range of theoretical interpretations. Namely, the  experimental constraints on dimension-6 operators should be reported as functions of the kinematic 
variables that set the relevant energy scale of the studied process.  This is especially important for hadron collider experiments where collisions probe a wide range of 
energy scales.    

\end{abstract}

\end{titlepage}
\newpage 

\renewcommand{\theequation}{\arabic{section}.\arabic{equation}} 

\section{Introduction}
\setcounter{equation}{0}

We consider an EFT where the SM is extended by a set of higher-dimensional operators, and assume that it reproduces the low-energy limit of a 
more fundamental UV description. The theory has the same field content and the same 
linearly-realized $SU(3) \times SU(2) \times U(1)$ local symmetry as the SM.
The difference is the presence  of operators  with canonical dimension $D$ larger than 4.  
These are organized in a systematic expansion in $D$, where each consecutive term is suppressed by a larger power of a high mass scale.
Assuming baryon and lepton number conservation, the Lagrangian takes the form~\cite{Buchmuller:1985jz,Hagiwara:1993ck,Grzadkowski:2010es}
\begin{equation}
\label{Lops}
{\cal L}_{\textrm{eff}}={\cal L}_{\rm SM}+ \sum_{i}  c_{i}^{(6)} {\cal O}_{i}^{(6)} \
+ \sum_{j}  c_{j}^{(8)} {\cal O}_{j}^{(8)}  + \cdots\,,
\end{equation}
where each ${\cal O}_i^{(D)}$ is a gauge-invariant operator of dimension $D$ and $c_i^{(D)}$ is the corresponding effective coefficient. 
Each coefficient has dimension $4-D$ and scales like a given power of the couplings of the UV theory; in particular, for an operator made of $n_i$ fields one has
\begin{equation}
\label{eq:counting}
c^{(D)}_i \sim \frac{(\text{coupling})^{n_i-2}}{(\text{high mass scale})^{D-4}}\, .
\end{equation}
This scaling holds in any UV completion which admits some perturbative expansion in its couplings.
It follows from simple dimensional analysis after restoring $\hbar\neq 1$ in the Lagrangian since couplings, as well as fields, carry $\hbar$ 
dimensions~\cite{Luty:1997fk,Cohen:1997rt,Giudice:2007fh} (see also Refs.~\cite{Pomarol:2014dya,Panico:2015jxa}).
An additional suppressing factor $(\text{coupling}/4\pi)^{2L}$ may arise with respect to the naive scaling if the operator is first generated at the $L^{\rm th}$-loop order
in the perturbative expansion.\footnote{See for instance Refs.~\cite{Arzt:1994gp,Einhorn:2013kja} for a discussion on whether a given operator can be generated 
at tree-level or at loop-level.} If no perturbative expansion is possible in the UV theory because it is maximally strongly coupled, 
then  Eq.~(\ref{eq:counting}) gives a correct estimate of the size of the effective coefficients by setting $\text{coupling}\sim 4\pi$.

The EFT defined by Eq.~(\ref{Lops}) is able to parametrize observable effects of a large class
of beyond the SM (BSM) theories. All decoupling BSM physics where new particles are much heavier than the SM ones and much heavier than the energy scale at which the experiment is performed can be mapped to such a Lagrangian. 
The main motivation to use this framework is that the constraints on the EFT parameters can be later re-interpreted as constraints on masses and couplings of 
new particles in many BSM theories. 
In other words, translation of experimental data into a theoretical framework has to be done only once in the EFT context, rather than for each BSM model separately. 
Moreover, the EFT can be used to establish a consistent picture of deviations from the SM by itself and thus can provide guidance for constructing a UV completion of the SM.

In the EFT, physical amplitudes in general grow with the energy scale of the process, due to the presence of  non-renormalizable operators.
Such framework  has therefore a limited energy range of validity.
In this note we address  the question of the validity range at the quantitative level 
{(similar questions have been addressed in Refs.~\cite{Berthier:2015oma,Berthier:2015gja,Greljo:2015sla} 
with partly different conclusions, and in Refs.~\cite{AguilarSaavedra:2011vw,Biekoetter:2014jwa,Englert:2014cva,Azatov:2015oxa,David:2015waa}; 
see also Refs.~\cite{delAguila:2008pw,delAguila:2010mx,Passarino:2012cb,Einhorn:2013kja,Gorbahn:2015gxa,deBlas:2014mba,Chiang:2015ura, Huo:2015exa,Huo:2015nka, Brehmer:2015rna, Wells:2015uba,Biekotter:2016ecg,Boggia:2016asg} for a discussion about matching UV models to the EFT, which indirectly addresses the question of its validity).}
We will discuss the following points: 
\bi 
\item 
Under what conditions does the EFT give  a faithful description of the low-energy phenomenology of some BSM theory? 
\item  When is it justified to truncate the EFT expansion at the level of dimension-6 operators?  To what extent {can}  experimental limits on dimension-6 operators 
be affected by the presence of dimension-8 operators?  Are there physically important examples where dimension-8 operators  cannot be neglected?
\item When is it justified to calculate the EFT predictions at tree level?  
In what circumstances may including 1-loop and/or real-emission corrections modify the predictions in a relevant way?   
\ei

It is important to realize that  addressing the above questions  
cannot be done in a completely model-independent way, but requires a number of (broad) assumptions about the new physics.
An illustrative example is that of the {\em Fermi theory}, 
which is an EFT for the SM degrees of freedom below the weak scale after the $W$ and $Z$ bosons have been integrated out. 
In this language, the weak interactions of the SM fermions are described at leading order by 4-fermion operators of $D$=6, such as: 
\beq
{\cal L}_{\rm eff} \supset 
 c^{(6)} \, (\bar e \gamma_\rho P_L \nu_e)    (\bar \nu_\mu \gamma_\rho P_L \mu) + {\rm h.c.} \, , 
 \qquad\quad c^{(6)} = -{g^2/2 \over m_W^2} = -{2 \over v^2}\, .
\eeq 
This operator captures several aspects of the low-energy phenomenology of the SM, including for example the  decay of the muon,  $\mu \to e \nu \bar\nu$,
and the inelastic scattering  of neutrinos on electrons $\nu e \to \nu \mu$.  
It can be used to adequately describe these processes as long as the energy scale involved ({\it i.e.} the momentum transfer between the electron current and the muon current)
is well below $m_W$.
However,  the information concerning $m_W$ is {\em not} available to a low-energy observer.
Instead, only the scale $|c^{(6)}|^{-1/2}  \sim  v = 2 m_W/g$ is measurable at low energies, and this is not sufficient to determine
$m_W$ without knowledge of the coupling~$g$.
For example, from a bottom-up viewpoint, a  precise measurement of the muon lifetime gives indications on the energy at which some new
particle ({\it i.e.} the $W$ boson) is expected to be produced in a higher-energy process, like
the scattering $\nu e \to \nu \mu$, only after making an assumption on the strength of its coupling 
to electrons and muons. 
Weaker couplings imply lower scales: for example, the Fermi theory could have ceased to be valid  right above the muon mass scale had
the SM been very weakly coupled,  $g \approx  10^{-3}$. On the other hand, a precise measurement of the muon lifetime sets an upper bound 
on the mass of the $W$ boson, $m_W \lesssim 1.5$~TeV,   
corresponding to the limit in which the UV completion is maximally strongly coupled, $g \sim 4 \pi$. 

This example illustrates the necessity of making assumptions -- in this case on the value of the coupling $g$ -- when assessing the validity range of the EFT, 
that is, when estimating the mass scale at which new particles appear (similar issues have been discussed in the context of Dark Matter searches, see for instance 
Refs.~\cite{Abdallah:2014hon,Racco:2015dxa}). On the other hand, the very interest in the EFT stems from its model-independence, and from the possibility of 
deriving the results from experimental analyses using Eq.~(\ref{Lops}) without any reference to specific UV completions.
In this note we identify under which physical conditions Eq.~(\ref{Lops}), and in particular its truncation at the level of dimension-6 operators, can be used 
to set limits on, or determine, the value of the effective coefficients.
Doing so, we also discuss the importance that results be reported by the experimental collaborations
in a way which makes it possible to later give a quantitative assessment of the validity range of the EFT approach used in the analysis.
As we will discuss below, this entails estimating the energy scale characterizing the physical process under study.
Practical suggestions on how experimental results could be reported will be given in this note.

\section{General discussion}
\setcounter{equation}{0} 
\label{sec:gen}

\subsection{Model-independent experimental results}

Let us first discuss how an experimental analysis can be performed  in the context of  EFT.
We start considering Eq.~(\ref{Lops}) truncated at the level of $D\!=\!6$ operators, and assume that it gives an approximate low-energy description of the UV theory. 
Further below we discuss the theoretical error associated with this truncation and identify the situations where the truncation is not even possible.
Physical observables are computed from the truncated EFT Lagrangian in a perturbative expansion according to the usual rules of effective field theories~\cite{Weinberg:1995mt}. The perturbative order to be reached depends on the experimental precision and on the aimed theoretical accuracy, as we
discuss in the following. Theoretical predictions obtained in this way are functions of the effective coefficients $c_i^{(6)}$ and can be used to perform a fit to 
the experimental data.
The impact of loop corrections on the fit can be estimated a posteriori based on the extracted values of (or limits on) the effective coefficients.
If some coefficients are smaller than others by as much as a loop factor $g_{SM}^2/16\pi^2$, where $g_{SM}$ is some SM coupling, then
1-loop corrections involving the larger coefficients might give a significant impact in their determination and should be included.
For example, it is well know that the operator correcting the top Yukawa coupling gives a 1-loop contribution to $h \to \gamma\gamma$
which must be included when computing this decay rate, since the sensitivity of the $t\bar{t}h$ measurement (which directly constrains the top Yukawa) 
is poorer than the one of the di-photon channel.
To the best of our knowledge, the only other two cases in which 1-loop insertions of \mbox{dimension-6} operators play a central role, given the current data from the LHC Run1, are the rates 
$gg\to h$ and $h\to Z\gamma$.
A more detailed discussion on the importance of 1-loop effects is given in Section~\ref{sec:loop}.
The fit to the coefficients  $c_i^{(6)}$ should be performed by correctly including the effect of all the theoretical uncertainties (such as those from the PDFs
and missing SM loop contributions\footnote{These latter can be estimated as usual by varying the factorization and renormalization scales.}) not originating 
from the EFT perturbative expansion. The errors due to the truncation at the $D\!=\!6$ level and higher-loop diagrams involving
insertions of different effective operators, on the other hand, are not quantifiable in a model-independent way and should thus be reported separately.
We will discuss how they can be estimated in the next sections.

Let us consider a situation in which no new physics effect is observed in future data (the discussion follows likewise in the case of observed deviations from the SM).
In this case, the experimental  results can be  expressed into the limits~\footnote{In general, the experimental constraints on different $c^{(6)}_i$ may have non-trivial 
correlations.  Depending on the chosen basis, the left-hand-side of  Eq.~(\ref{eq:bounds}) may contain linear combinations of several effective coefficients.
If a deviation from the SM is observed, Eq.~(\ref{eq:bounds}) turns into a confidence interval, 
$\delta_i^\text{d,exp}(M_{\rm cut}) <  c^{(6)}_i < \delta_i^\text{u,exp}(M_{\rm cut})$.}
\begin{equation}
\label{eq:bounds}
c^{(6)}_i < \delta_i^\text{exp}(M_{\rm cut})\,.
\end{equation}
The functions $\delta_i^\text{exp}$ depend on the {upper} value,
here collectively denoted by $M_{\rm cut}$,  of the kinematic variables (such as transverse momenta or 
invariant masses) that set the typical energy scale characterizing the process. 
In general, Eq.~(\ref{eq:bounds}) is obtained by 
imposing cuts on these variables and making use of the differential kinematic distributions of the process.

There are situations in which the relevant energy of the process is fixed by the kinematics.
For example, in inclusive on-shell Higgs decays  one has $M_{\rm cut} \approx m_h$. 
Another example is  $e^+ e^- $ collisions at a fixed center-of-mass energy $\sqrt{s}$, in which case $M_{\rm cut} \approx \sqrt{s}$. 
On the other hand, the relevant scale for the production of two on-shell particles  in proton-proton collisions
is the center-of-mass energy of the partonic collision $\sqrt{\hat s}$, which varies in each event and may not be fully reconstructed in practice.
Important examples of this kind are vector boson scattering processes  ({\it e.g.} with final states $WW\to 2l2\nu$ and $ZZ\to 4l$), and Higgs  production in association with 
a vector boson ($Vh$) or a jet ($hj$). In all these processes the relevant energy scale is given by the invariant mass of the final pair. 
Since the energy scale of the process determines the range of validity of the EFT description, it is extremely important that the experimental limits $\delta_i^\text{exp}$  
are reported by the collaborations for various values of~$M_{\rm cut}$.
For processes occurring over a wide energy range (unlike Higgs decays or $e^+ e^- $ collisions),  knowledge of {only the limit $\delta_i^\text{exp}$
obtained by making use of all the events without any restriction on the energy  ({\it i.e.} for $M_\text{cut}\to \infty$)}
severely limits the interpretation of the EFT results in terms of constraints on specific BSM models. 

If the relevant energy of the process cannot be determined, because for example the kinematics cannot be closed, then setting consistent bounds 
requires a more careful procedure,  {for example similar to the one proposed in Ref.\cite{Racco:2015dxa} in the context of DM searches}. 
In these cases other correlated (though not equivalent) variables may be considered, as for example the transverse momentum of the Higgs boson or a lepton
in $Vh$ production.

\subsection{EFT validity and interpretation of the results}

Extracting bounds on {(or measuring)} the EFT coefficients can be done by experimental collaborations
 in a completely model-independent way. 
However, the {\em interpretation} of these bounds  is always model-dependent. 
In particular, whether or not  the EFT is valid in  the parameter space probed by the experiment  
depends on  further assumptions about the (unknown) UV theory. 
These assumptions {correspond, in the EFT language, to a choice of power counting},
{\it i.e.} a set of rules to estimate the  coefficients of the effective operators in terms of the couplings and mass scales of the UV dynamics. 

The simplest situation, which we discuss in some detail here, is when the microscopic dynamics is characterized by a single mass scale $\Lambda$ and a single 
new coupling $g_*$~\cite{Giudice:2007fh}. 
This particular power-counting  prescription smoothly interpolates between the naive dimensional analysis ($g_* \sim 4 \pi$)~\cite{Manohar:1983md,Cohen:1997rt}, 
the case $g_* \sim  1$ as for example in the Fermi theory, and the very weak coupling limit $g_* \ll 1$.    
While this is not a unique prescription, it covers a large selection of popular scenarios beyond the SM.
In this class falls the Fermi theory described previously, as well as other weakly-coupled models where a narrow resonance 
is integrated out. 
Moreover, despite the large number of resonances, also some theories with a  strongly-interacting BSM sector belong to this category ({\it e.g.} the holographic composite 
Higgs models~\cite{Agashe:2004rs} or, more generally, theories where the strong sector has a large-$N$ description).  
The scaling of the effective coefficients with $g_*$ is determined by Eq.~(\ref{eq:counting}) and by symmetries and selection rules.

For example,  if the coupling strength of the Higgs field to the new dynamics is $g_*$, then the coefficient of an operator with four Higgs fields and two derivatives 
scales like $g_*^2$ (see Table~\ref{tab:ops}). Approximate symmetries acting in the low-energy theory can reduce the maximal scaling with $g_*$ of the coefficients.
For instance, approximate chiral symmetry implies that the coefficient of an operator with a fermion scalar bilinear and three Higgs fields scales  as $y_f g_*^2$,  
where $y_f$ is the corresponding Yukawa coupling. Some examples relevant for Higgs physics are reported in Table~\ref{tab:ops} 
{(for examples of alternative power-counting  and selection rules schemes see Refs.~\cite{Pomarol:2014dya,Azatov:2015oxa,Liu:2016idz}).}
As a final illustrative case consider the complete Fermi theory, where the approximate flavor symmetry of the SM is inherited by the low-energy EFT, entailing 
a suppression of flavor-violating 4-fermion operators.
%
\begin{table}[!tb]
\renewcommand{\arraystretch}{1.3}
\begin{center}
\small
\begin{tabular}{|l|c|l|}
\hline
Operator & Naive (maximal) & Symmetry/Selection Rule \\[-0.17cm]
& scaling with $g_*$ & and corresponding suppression \\[0.1cm]
\hline
$O_{y_\psi} = |H|^2    \bar \psi_L H \psi_R$ & $g_*^3$& Chiral: $y_f/g_*$\\[0.1cm]
\hline && \\[-0.55cm]
$O_T =(1/2) \left(H^\dagger {\lra{D}_\mu} H\right)^2$& $g_*^2$& Custodial: $(g^\prime/g_*)^2$, $y_t^2/16\pi^2$\\[0.25cm]
\hline
\multirow{4}{*}{\hspace{-2mm}\begin{tabular}{l} $O_{GG} = |H|^2 G^a_{\mu\nu}G^{a\,\mu\nu}$ \\[0.12cm] $O_{BB} = |H|^2 B_{\mu\nu}B^{\mu\nu}$\end{tabular}} 
   & \multirow{4}{*}{$g_*^2$} & Shift symmetry: $(y_t/g_*)^2$ \\
& & Elementary Vectors: $(g_s/g_*)^2$ (for $O_{GG}$) \\[-0.09cm]
& & \hspace{3.45cm} $(g'/g_*)^2$ (for $O_{BB}$) \\[0.09cm]
& & Minimal Coupling: $g_*^2/16\pi^2$ \\[0.1cm]
\hline
$O_6 = |H|^6$ & $g_*^4$ & Shift symmetry: $\lambda/g_*^2$\\[0.1cm]
\hline
$O_H = (1/2) (\partial^\mu |H|^2)^2$ & $g_*^2$& Coset Curvature: $\epsilon_c$\\[0.1cm]
\hline && \\[-0.55cm]
$O_B = (i/2)\left( H^\dagger  \lra {D^\mu} H \right )\partial^\nu  B_{\mu \nu}$ & \multirow{2}{*}{$g_*$} & 
\multirow{2}{*}[1.3mm]{\hspace{-2mm}\begin{tabular}{l} Elementary Vectors: $g'/g_*$ (for $O_B$) \\[-0.09cm]\hspace{3.45cm} $g/g_*$ (for $O_W$) \end{tabular} }\\[0.2cm]
$O_W = (i/2) \left( H^\dagger \sigma^a \lra {D^\mu} H \right )\partial^\nu  W^a_{\mu \nu}$ & & \\[0.25cm]
\hline
\multirow{3}{*}{\hspace{-2mm}\begin{tabular}{l} $O_{HB} = (i/2) \left( D^\mu H^\dagger   {D^\nu} H \right )  B_{\mu \nu}$ \\[0.12cm]
                           $O_{HW} = (i/2) \left( D^\mu H^\dagger \sigma^a  {D^\nu} H \right )  W^a_{\mu \nu}$ \end{tabular}}
 & \multirow{3}{*}{$g_*$} & Elementary Vectors: $g'/g_*$ (for $O_{HB}$) \\[-0.1cm]
& & \hspace{3.45cm}  $g/g_*$ (for $O_{HW}$) \\[0.1cm]
& & Minimal Coupling: $g_*^2/16\pi^2$\\[0.1cm]
 \hline
\end{tabular}
\end{center}
\caption{\emph{Some operators relevant for Higgs physics and the impact of approximate symmetries on the estimated size of their coefficient~\cite{Giudice:2007fh}. 
The coefficient $\epsilon_c$ parametrizes the possibility that the Higgs doublet originates as a PNGB from the flat coset $ISO(4)/SO(4)$~\cite{Liu:2016idz} 
(see also~\cite{Alonso:2015fsp}). A suppression $g_V/g_*$ for every field strength (referred to as Elementary Vectors in the table), applies to all models where the 
transverse components of gauge bosons are elementary. See Ref.~\cite{Liu:2016idz} for a construction where transverse gauge bosons are composite and have strong 
dipole interactions.
}
}\label{tab:ops}
\end{table}
%

For a given power counting, it is relatively simple to derive limits on the theoretical parameter space that are automatically consistent with the EFT expansion, provided 
the relevant energy of the process is known. Consider  the case of a single scale $\Lambda$ and a single coupling strength~$g_*$. 
Then the bounds~(\ref{eq:bounds}) can be recast as limits on these two parameters by using the power counting to estimate $c_i^{(6)} = \tilde c^{(6)}_i(g_*)/\Lambda^2$, 
and setting the {upper value of} the relevant energy scale to $M_\text{cut} = \kappa \Lambda$. 
Here $\tilde c_i^{(6)}(g_*)$ is a (dimensionless) polynomial of $g_*$ and of the SM couplings, while $0 < \kappa < 1$ controls the size of the tolerated error due to neglecting 
higher-derivative operators (the value of $\kappa$ can be chosen according to the sensitivity required in the analysis).
One finds
\begin{equation} \label{eq:bounds2}
\frac{\tilde c^{(6)}_i(g_*)}{\Lambda^2} < \delta_i^\text{exp}(\kappa\Lambda)   \, .
\end{equation}
These inequalities determine the region of the plane $(\Lambda, g_*)$ which is excluded consistently with the EFT expansion for a given $\kappa$.
This is a conservative bound, since it is obtained by using only a subset of the events (effectively only those with relevant energy up to 
$M_\text{cut} = \kappa \Lambda$).
It is thus less stringent than the bound one would obtain in the full theory with the full dataset, but it is by construction consistent with the EFT expansion.
Compared to the constraint implied by the full theory with the same reduced dataset, that of Eq.~(\ref{eq:bounds2}) has an error of order~$\kappa^2$.
For constraints obtained in this way, and for a valid EFT description in general, no question of unitarity violation arises (see for example 
Ref.~\cite{Degrande:2012wf} for a discussion of this issue in the context of anomalous triple gauge couplings).

An analysis of the experimental results based on the multiple cut technique proposed here was performed in Ref.~\cite{Biekoetter:2014jwa} 
for the $Vh$ associated production. The same strategy can  also be  applied to more complicated theories following a  power counting different than the simple 
$g_*$-scaling discussed above (see for example Ref.~\cite{Azatov:2015oxa}).

\subsection{On the necessity of a power counting}\label{sec:PC}

The necessity of an appropriate power counting stems from a number of reasons. First of all it provides a physically motivated range in which the coefficients 
$c_i^{(D)}$ are expected to vary. 
Secondly, and very importantly, it allows one to estimate the relative importance of higher-order terms in the EFT series.
If two operators with dimension $D$ and $D+2$ contribute at  tree-level to the same vertex, 
then they must have the same field content after electroweak symmetry breaking.
In this case the higher-dimensional operator must have two more powers of the Higgs
field or two more derivatives compared to the first operator. 
Its contribution to the vertex is thus naively suppressed by a  factor equal to, respectively, 
\begin{equation}
\label{eq:D8scal}
\kappa_v^2 = \left(\frac{g_* v}{\Lambda}\right)^2\quad \textrm{and}\quad  \kappa_E^2 = \left(\frac{E}{\Lambda}\right)^2,
\end{equation}
where $E \approx M_{\rm cut}$ is the (maximum) energy characterizing the process 
under consideration (hence $\kappa_E \approx \kappa$).
The EFT series is  built in terms of these two expansion parameters, which must be both small for the description to be valid.

From Eq. (\ref{eq:D8scal}),  one would  expect  the contribution of higher-dimensional operators to a given observable to be of relative order 
$\kappa_v^2$ or $\kappa_E^2$, hence
always subdominant  compared to the contribution of lower-dimensional operators, in the case of a valid EFT expansion. 
This naive estimate however assumes that the hierarchy in the effective coefficients is entirely dictated by their scaling with $\Lambda$,
{\it i.e.} that the dimensionless coefficients $\tilde c_i^{(D)} \equiv c_i^{(D)} \Lambda^{D-4}$ 
are all characterized by the same underlying interaction strength.
It may happen, on the other hand, that a stronger interaction strength only appears at a higher level in the perturbative expansion as the result
of some selection rule or symmetry. In this case the next-to-leading correction from higher-dimensional operators might become sizable and
even dominate over the naively leading contribution. 

As an example illustrating the above possibility consider a $2\to 2$ scattering process, where the SM contribution to the amplitude is at most of order $g_{SM}^2$ 
at high energy ($g_{SM}$ denotes a SM coupling). 
The correction from $D\!=\!6$ operators involving derivatives will in general grow quadratically with the energy and can be as large as 
$g_*^2 (E^2/\Lambda^2)$.~\footnote{Effects growing with energy can also be induced by operators without additional derivatives, if they yield new contact interactions relevant for the process, or if they disrupt cancellations between energy-growing individual contributions of different SM diagrams, 
see {\it e.g.}~\cite{Chanowitz:1985hj,Appelquist:1987cf,Maltoni:2001hu,Contino:2010mh,Dror:2015nkp,Bylund:2016phk}.}
If the coupling strength $g_*$ is much larger than $g_{SM}$, then the BSM contribution dominates over the SM one at sufficiently high energy 
({\it i.e.} for $\Lambda > E > \Lambda\, (g_{SM}/g_*)$), while the EFT expansion is still valid. 
The largest contribution to the cross section in this case comes from the square of the $D\!=\!6$ term, rather than from its interference with the SM.
The best sensitivity to $c_i^{(6)}$ is thus expected to come from the highest value of the relevant energy scale accessible in the experiment.
In this example the contribution of $D\!=\!6$ derivative operators is 
enhanced by a factor $(g_*/g_{SM})^2$ compared to the naive expansion parameter~$\kappa_E^2$; such enhancement is a consequence of the fact that 
the underlying strong coupling $g_*$ only appears at the level of $D\!=\!6$ operators, while {$D\!=\!4$} operators mediate weaker interactions.
Here, no further enhancement exists between $D\!=\!6$ and $D\!=\!8$ operators, {\it i.e.} $D\!=\!8$ operators are subdominant and the EFT series is converging.
In other words, although the contributions to the cross section proportional to $(c_i^{(6)})^2$ and $c_i^{(8)}$ 
are both of order $1/\Lambda^4$, the latter (generated by the interference of $D\!=\!8$ operators with the SM) is smaller by a factor
$(g_{SM}/g_*)^2$  independently of the energy, and can thus be safely neglected.  
A well known process where the above situation occurs
 is the scattering of longitudinally-polarized vector bosons.
 Depending on the UV dynamics, the same can happen in other $2\to 2$ scatterings,
such as Higgs associated production with a $W$ or $Z$ boson (VH)~\cite{Biekoetter:2014jwa,Biekotter:2016ecg}, dijet searches at the LHC~\cite{Domenech:2012ai} or top physics~\cite{Dror:2015nkp,AguilarSaavedra:2011vw}.

{Another situation in which  $(c_i^{(6)})^2$ terms dominate is when the SM interactions are suppressed by some accidental (possibly approximate) symmetry
not respected by the BSM dynamics. Consider for example the corrections to flavor-changing neutral current processes (strongly suppressed 
in the SM by a loop and CKM factors), that would originate from BSM theories that are not Minimal Flavor Violating (see {\it e.g.}~\cite{Durieux:2014xla}).
An even sharper example is lepton-flavor violating processes ({\it e.g.} $h \to \mu \tau$), for which the SM amplitude exactly vanishes.

The examples discussed show that, for structural reasons, a stronger interaction may be revealed at the $D$=6 level in the EFT expansion.
It is also possible, on the other hand, that such stronger interaction appears only at the level of $D\!=\!8$ operators, so that these dominate over $D$=6 ones and 
over the SM  in the high-energy regime. In this case  the $D$=6 EFT description may be inadequate, as we discuss in Section~\ref{sec:lim}.}

As a final remark on the importance of higher-order operators, notice that the bounds of Eq.~(\ref{eq:bounds2}) can also be interpreted with a different perspective. 
Rather than specifying an error tolerance $\kappa$ and  extracting information on $g_*$ and~$\Lambda$, one can make BSM assumptions on either $g_*$ or 
$\Lambda$ and see to what precision they can be measured. 
For instance, consider the  case in which the same coupling strength $g_*$ controls the size of all the effective coefficients~$\tilde c_i^{(D)}$.
Then, from Eqs.~(\ref{eq:bounds}) and (\ref{eq:D8scal}) it follows that  the uncertainty due to neglecting $D\geq 8$ operators can be expressed 
in terms of the experimental accuracy $\delta^{\rm exp}$ on the bounds (or measurements) of the effective coefficients as follows:
\begin{equation}
\label{eq:errF}
\kappa_E^2 \lesssim
\frac{E^2}{g_*^2}\,\delta^{\rm exp}(E) \,,\quad \kappa_v^2 \lesssim
v^2\, \delta^{\rm exp}(E)\, .
\end{equation}
From this expression it becomes clear that, for a given experimental precision and energy, BSM theories with larger $g_*$ will be constrained with better accuracy. 
In fact, Eq.~(\ref{eq:errF}) can be used to estimate the experimental precision needed to constrain a coefficient
$c^{(6)}$ in a meaningful way for a given $E=M_{\rm cut}$, within the validity of the EFT. 
Since the $\kappa_i$ are bounded from above by these equations, one can explicitly derive the value of $\delta^{\rm exp}$
that guarantees $\kappa_i<1$.

\subsection{On the importance of loop corrections}
\label{sec:loop}

So far our discussion was limited to tree-level effects of $D\!=\!6$ operators. 
The EFT can be consistently extended to an arbitrary loop order 
by computing observables perturbatively in the SM couplings.
The corresponding series is controlled by the expansion parameter $g_{SM}^2/16 \pi^2$, which adds to the two EFT parameters $\kappa_v^2$
and $\kappa_E^2$ already discussed.
One-loop effects of $D$=6 operators are formally suppressed by $O(g_{SM}^2/16 \pi^2)$, and are thus {generally} subleading compared to the tree-level
contributions.
Including loop corrections in the EFT context is, at present, less crucial than for a pure SM calculation. 
This is because the experimental precision is typically better than the magnitude of the SM loop corrections, therefore going beyond tree level
in a SM calculation  is essential to obtain a correct description of physical processes.
In the case of the EFT, on the other hand, we are yet to observe any leading-order effect of higher-dimensional operators. 

There do  exist situations, however,  where including NLO corrections may be important for obtaining an adequate description of physical processes in the EFT  (see Refs.~\cite{Willenbrock:2014bja,Henning:2014wua} for an extended discussion). 
For example, it is well known that  NLO QCD corrections to the SM predictions of certain processes
at the LHC can be of order 1, and  large k-factors are expected to apply to the EFT corrections as well.
Another example is the one-loop Higgs corrections to electroweak precision observables.
Since deviations of the Higgs couplings due to $D$=6 operators can be relatively large (up to $O(10\%)$) without conflicting with current experimental data, 
the 1-loop effects, in spite of the suppression factor,   can be numerically important for observables measured with a per-mille 
precision~\cite{Barbieri:2007bh,Elias-Miro:2013eta,Henning:2014gca}.
Similarly, four-fermion operators can contribute at one loop to the Higgs decays $h\to b\bar b$ and $h\to\tau \bar \tau$
and be effectively constrained by these processes, as recently pointed out in Ref.~\cite{Gauld:2015lmb}.
Along the same lines, competitive indirect bounds on CP-violating operators can be
obtained by considering their loop corrections (including operator mixing via RG evolution) to well-measured 
electromagnetic dipole observables \cite{Kamenik:2011dk,McKeen:2012av,Brod:2013cka,Cirigliano:2016njn}.

More generally, 1-loop corrections are important if they stem from large coefficients and correct precisely measured observables whose tree-level contribution
arises from smaller coefficients.
The tree-level contribution of a $D\!=\!6$ operator may be suppressed, for example, because its coefficient is
generated  at the 1-loop level by the UV dynamics.
In this case, both the 1-loop and tree-level
contributions from $D\!=\!6$ operators would correspond to 1-loop processes in the UV theory.
An example of this kind is the decay of the Higgs boson to two photons, $h \to \gamma \gamma$, which arises necessarily at the 1-loop level if the UV theory
is minimally coupled (see Ref.~\cite{Giudice:2007fh} and the appendix of Ref.~\cite{Liu:2016idz}) and perturbative.
It is interesting to notice that the bulk of the 1-loop corrections from $D\!=\!6$ operators corresponds to the RG evolution of their coefficients~\cite{Ghezzi:2015vva,Hartmann:2015oia, Hartmann:2015aia}.
The remaining finite (threshold) corrections are instead usually smaller since they bear no logarithmic enhancement. 
Performing a fit in terms of the coefficients evaluated at the low-energy scale thus automatically re-sums their RG running from the new physics scale, hence
the bulk of the 1-loop corrections. In this sense, as long as finite terms can be neglected, an explicit evaluation of the 1-loop insertions of $D\!=\!6$ operators is required  
only if the observables included in the fit are characterized by widely different energy scales, or if one wants to match to the UV theory at the high scale.
The calculation of NLO effects in the context of the EFT is currently an active field of study, 
see for instance the recent papers  in the last 15~months\cite{Dawson:2015gka,Franzosi:2015osa,Drozd:2015kva,Grober:2015cwa,Hartmann:2015oia,Ghezzi:2015vva,Hartmann:2015aia,David:2015waa,Grazzini:2015gdl,Gauld:2015lmb,Mimasu:2015nqa,Drozd:2015rsp,Wells:2015cre,Zhang:2016snc,Zhang:2016omx,delAguila:2016zcb,Grober:2016wmf,Henning:2016lyp,Ellis:2016enq}.
As suggested by the above discussion, it is very important to identify all cases  where 1-loop effects of $D\!=\!6$ operators can be relevant.

Besides one-loop effects, it is sometimes also important to include corrections from real emission processes.
In particular, including additional jets may be important when exclusive observables, {\it i.e.} quantities particularly sensitive to extra radiation,
are studied. An example is given by the transverse momentum distribution of leptons in the process $pp \to h \to VV \to 4\ell$, for which NLO real emissions
are known to give $O(1)$ effects (see {\it e.g.}~\cite{Sapeta:2013vaa,Grazzini:2008tf}).

\vspace{0.4cm}
To summarize, in this section we have discussed how  using a power counting is required to assess the validity of an EFT truncated at the level of $D\!=\!6$ 
operators and interpret the experimental results in terms of physical masses and couplings of the UV theory. From a practical point of view, the power counting 
allows one to estimate the relative importance of $D\!=\!6$ and $D\!=\!8$ operators (for instance deducing  when ${\tilde c}_i^{(6)}\simeq {\tilde c}_i^{(8)}$).
This framework is particularly well suited to interpret the Higgs  data at the LHC.
One important observation from this discussion is that although the cut-off scale is an integral part of the EFT formulation, 
its value cannot  be directly determined  from  low-energy experiments. 
In order to estimate its value and the range of validity of the EFT, results should be presented by the experimental collaborations as a function of 
{the upper bounds, here collectively denoted with $M_{\rm cut}$,  on} the kinematic variables that set the relevant energy of the process.  
For the purpose of estimating the validity of the EFT approach, it might be useful to compare the constraints obtained with 
and without including the quadratic contributions of $D\!=\!6$ operators in the theoretical calculations of observables: significant differences between these two 
procedures
will indicate that the results apply only in the case of strongly-coupled UV theories, where quadratic terms can give the dominant effect at large energies.
Finally, notice that even in situations where it makes sense to expand the cross section at linear order in the coefficients of  $D\!=\!6$ operators, quadratic terms 
should always be retained in the calculation of the likelihood function, as we discuss in detail in the Appendix.

\section{Limitations of the $D\!=\! 6$ EFT}
\setcounter{equation}{0} 
\label{sec:lim}

The SM Lagrangian extended by $D\!=\! 6$ operators is an effective theory that captures the low-energy regime of a large class of models with new heavy particles.
However, not every such model can be adequately approximated by truncating the EFT expansion at the $D\!=\! 6$ level.  
In this section we discuss these special cases where a more complicated approach is required.

As argued in Section~\ref{sec:gen}, generically one expects that the effect of  $D\!=\! 8$ operators is subleading compared to that of  $D\!=\! 6$ ones at energies 
$E\ll \Lambda$,  with $\Lambda \gg m_W$. 
On the other hand, if $E \sim \Lambda$, the entire tower of operators ($D\!=\! 8$, $D\!=\! 10$, etc.) contributes, and  the EFT expansion is not useful. 
Nevertheless, there are physical situations when $D\!=\! 8$ operators can be relevant,   
despite the whole EFT expansion being convergent. 
We identify the following cases: 
\bi

\item 
{\bf Symmetries} \\
As previously noticed, the contribution of $D\!=\! 8$ operators can be dominant if they mediate a strong interaction that does not appear at lower level in the EFT 
expansion.  This happens if interactions generated by $D\!=\! 6$ and $D\!=\! 4$ operators remain weaker as a consequence of some (approximate) symmetry 
of the low-energy theory.
An example of such situation occurs in  models with a pseudo Nambu--Goldstone boson Higgs, where $D\!=\! 6$ and $D\!=\! 8$ operators contributing to 
Higgs pair production via gluon fusion are generated by different mechanisms~\cite{Azatov:2015oxa}.
The  $D\!=\! 6$ operator $|H|^2G^a_{\mu\nu}G^{a,\mu\nu}$ is not invariant under the Higgs  shift symmetry (which is part of the Goldstone symmetry) and its coefficient
is proportional to the square of some small coupling that breaks it, see Table~\ref{tab:ops}.
On the other hand,   two $D$=8 operators with extra derivatives can be constructed 
{($D_\lambda H^\dagger D^\lambda H G^a_{\mu\nu}G^{a,\mu\nu}$ and $D^\mu H^\dagger D^\nu H G^a_{\mu\alpha}G^{a,\alpha}_\nu$)}
that respect the shift-symmetry and whose coefficients are therefore  unsuppressed. 
As a consequence, in the energy range $\Lambda \sqrt{\tilde c^{(6)}/\tilde c^{(8)}} < E < \Lambda$ the contribution from $D\!=\! 8$ operators dominates over that from 
$D$=6 ones but the EFT expansion is still valid.
Another example is given by theories where dipole interactions of $SU(2)_L$ gauge bosons ({\it i.e.} those involving the field strength) are associated with a 
new strong coupling, while monopole interactions generated by covariant derivatives are weak~\cite{Liu:2016idz}. The protecting symmetry in this case is a global
$SU(2)_L^{global}\times U(1)_{local}^3$ (as opposed to the local $SU(2)_L$) which is obtained in the limit of vanishing weak gauge coupling.
In these theories $D\!=\! 8$ operators give the leading contribution to scattering processes where the contribution from $D\!=\! 6$ operators involve some weak 
coupling.
For instance, the leading contribution to the scattering of transversely-polarized vector bosons comes from the dimension-8 operator~$W_{\mu\nu}^4$. 
Similar conclusions also hold for fermions if they are identified with the goldstinos of $N$ spontaneously-broken 
symmetries~\cite{Bardeen:1981df}. In this case the first interactions respecting supersymmetry arise at dimension 8 and include self interactions
of the form $\bar\psi^2\partial^2\psi^2$~\cite{Volkov:1973ix}.

\item{\bf Zero at leading order} \\ 
For certain processes,  contributions to the scattering amplitude from $D\!=\! 6$ operators vanish without any symmetry reason.   
If that is the case, the first non-trivial corrections appear only at the $D\!=\! 8$ level.  
One well known example is the s-channel production of neutral gauge boson pairs.  Such a process does not occur in the SM nor in the 
$D\!=\! 6$ EFT  because triple gauge couplings of neutral gauge bosons arise only from $D \geq 8$ operators~\cite{Degrande:2013kka} {(similar issues arises with certain combinations of charged triple gauge bosons~\cite{Arzt:1994gp})}. 
This category includes also  $2\to 2$ scattering processes involving transverse gauge bosons,  where, because of the helicity structure of the amplitudes,  
the dimension-6 operators do not interfere with the SM while the dimension-8 ones do~\cite{coninoprop}. 
Another example is the triple Higgs production by vector boson fusion whose energy-growing piece originates from $D \geq 8$ operators only~\cite{Belyaev:2012bm,Contino:2013gna}.

 \item {\bf Selection Rules inherited from the UV dynamics} \\ 
Approximate selection rules enhancing the contribution of $D\!=\! 8$ operators relative to $D\!=\! 6$ ones can arise in the low-energy theory as the consequence 
of symmetries  (not acting on low-energy fields) or structural features of the UV dynamics. 
An example of this kind is given by a 5-dimensional theory with a light Kaluza-Klein (KK) graviton or radion,
or, similarly, a 4-dimensional theory with a light dilaton.
The KK graviton, the radion and the dilaton all couple to the stress-energy tensor, which has canonical dimension 4.
The leading low-energy effects at tree level
are thus encoded in $D\!=\! 8$ operators, while $D\!=\! 6$ ones are generated only at the loop level.

\item {\bf Fine-tuning} \\ 
One can imagine a fine-tuned situation where integrating out  the heavy states in the UV theory generates $D\!=\! 6$ operators with coefficients 
that are accidentally much smaller than their naive estimate and much smaller than those of the $D\!=\! 8$ operators, $\tilde c_i^{(6)} \ll \tilde c_i^{(8)}$.
In such a case, the EFT with only $D\!=\! 6$ operators will not correctly approximate the dynamics of the UV theory.
By nature, naive dimensional analysis and simple power counting are just not suited when some parameters are accidentally small.
\ei 

Notice that, contrary to the structural hierarchies 
described in the first two points,  those from fine tuning or UV selection rules 
are in general not stable under Renormalization Group evolution in the UV theory. There might be special situation, however, in which the hierarchy in
the coefficients at the matching scale is not spoiled by running down to low energies~\footnote{{As an example consider the non-renormalization  theorems~\cite{Elias-Miro:2014eia,Cheung:2015aba} that lead to a specific structure of the 1-loop anomalous dimension matrix for dimension-6 operators~\cite{Grojean:2013kd,Elias-Miro:2013gya,Elias-Miro:2013mua, Jenkins:2013zja, Jenkins:2013wua, Alonso:2013hga,Elias-Miro:2013eta,Alonso:2014rga}.}}.

If any of the above mechanisms is at work, $D\!=\! 8$ operators can give the leading correction to a given observable and should  be included in the EFT description.
Moreover, since the present experimental constraints on physics beyond the SM display a hierarchical structure (for example,  electroweak precision observables 
were measured by LEP-1 with  per-mille accuracy,  while LHC Higgs observables are currently measured with an $O(10\%)$ accuracy at best), 
in some explicit scenarios $D\!=\! 8$ operators may be phenomenologically as important as $D\!=\! 6$ ones to obtain information on the mass scale 
and the couplings of the UV theory (complete classifications of $D$=8 operators have recently appeared in the literature, see 
Refs.~\cite{Henning:2015daa,Lehman:2015coa,Henning:2015alf}).

In summary, there do exist physical situations where the inclusion of dimension-8 operators (on top of or instead of dimension-6 ones) is well motivated and where, 
nonetheless,   the EFT expansion remains well defined. 
 This does not mean, however, that introducing a complete set  of $D\!=\! 8$ operators into EFT analyses is preferable in general. 
Such a framework would be utterly complicated,  
and, moreover,  the existing experimental data do not contain enough information  to lift the degeneracy between  $D\!=\! 6$ and $D\!=\! 8$ operators. 
Instead, it is suggested to focus on the (already challenging) EFT with $D\!=\! 6$ operators and address case-by-case the special situations discussed above.    
Nevertheless, it is important to be aware of such special situations where a $D\!=\! 6$ truncated EFT fails.
{In these cases a consistent EFT description must obviously include dimension-8 operators, since these give the leading contribution.}

\section{An Explicit Example}\label{app:example}
\label{app:BSMmodel}

In this section we illustrate our general arguments by comparing the predictions of the EFT and of a specific BSM model which reduces to that EFT at low energies. 
To this end we discuss the $q \bar q \to V h$ process at the LHC, along the lines of Ref.~\cite{Biekoetter:2014jwa}.  
The purpose of the example presented below is to demonstrate that, as in the Fermi theory, the knowledge of the $D\!=\!6$ coefficients of an effective Lagrangian is not enough to determine  the validity range of the EFT approximation. 
Therefore, the theoretical error incurred as a result of the truncation of the EFT Lagrangian cannot be quantified in a model-independent way.

We consider the SM extended by a triplet of vector bosons $V_\mu^i$ with mass $M_V$  transforming in the adjoint representation  of the SM $SU(2)_L$ symmetry. 
Its couplings to the SM fields are described by~\cite{Low:2009di,deBlas:2012qp,Pappadopulo:2014qza}
\beq
\label{eq:lvt}
{\cal L} \supset    i g_H  V_\mu^i H^\dagger \sigma^i  \overleftrightarrow{D_\mu} H 
+   g_q  V_\mu^i  \bar q_L \gamma_\mu \sigma^i q_L,
\eeq 
where $q_L = (u_L,d_L)$ is a doublet of the 1st generation left-handed quarks. 
In this model $V_\mu^i$ couples to light quarks, the Higgs boson, and electroweak gauge bosons, 
and it contributes to the $q \bar q \to V h$ process at the LHC. 
Below the scale $M_V$,  the vector resonances can be integrated out, giving rise to an EFT where the SM is extended by $D\!=\!6$ and higher-dimensional operators. 
Thus, $M_V$ plays the role of the EFT cut-off scale $\Lambda$. 
Using the language of the Higgs basis~\cite{HB}, at the $D\!=\!6$ level the EFT is described by the parameter  $\delta c_z$  (relative correction to the SM Higgs couplings to $WW$ and $ZZ$) and $\delta g_L^{Zq}$ (relative corrections to the $Z$ and $W$ boson couplings to left-handed quarks),   plus other parameters that do not affect the $q \bar q \to V h$ process at tree level. 
The relevant EFT parameters are matched to those in the UV model as 
\beq 
\label{eq:eftvsmodel}
\delta c_z  =  -  {3 v^2 \over 2 M_V^2} g_H^2, 
 \qquad  
\, [\delta g^{Zu}_L]_{11}  = - [\delta g^{Zd}_L]_{11}  =   - {v^2 \over 2 M_V^2}  g_H g_{q}\, . 
 \eeq 
When these parameters are non-zero,  the EFT amplitude for the scattering $q \bar q \to V h$ with a longitudinal vector boson grows as the square of the partonic
center-of-mass energy $s \equiv M_{Wh}^2$ (the transverse amplitude grows instead linearly with the energy).
Then, for a given value of the EFT parameters, the deviation from the SM prediction becomes larger.
However, above a certain energy scale, the EFT may no longer approximate correctly the UV theory defined by Eq.~(\ref{eq:lvt}), and as such an experimental constraint 
on the EFT parameters does not provide any information about the UV theory. 
  
\begin{figure}[t]
\centering
\includegraphics[width=0.49 \textwidth]{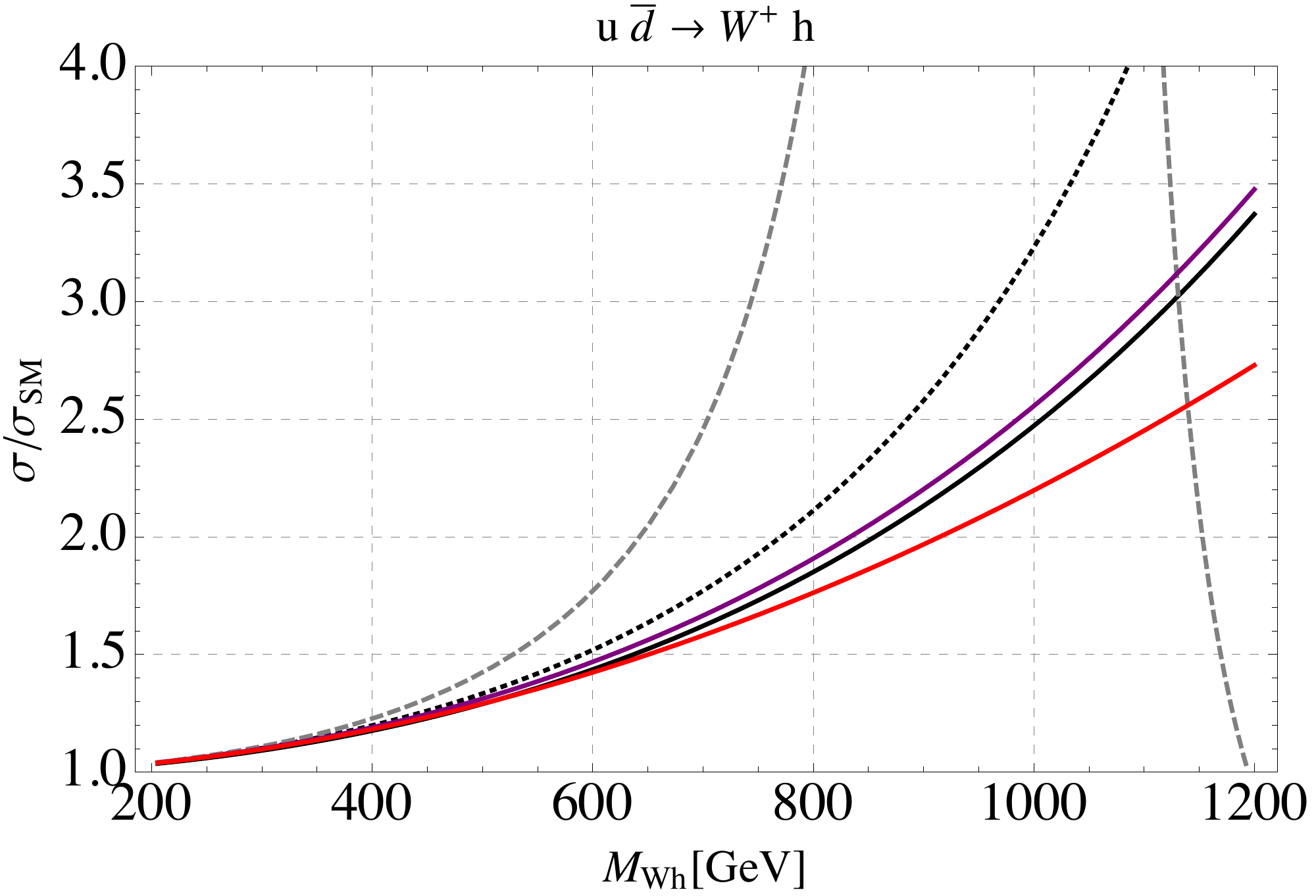}
\hspace{0.05cm} 
\includegraphics[width=0.49 \textwidth]{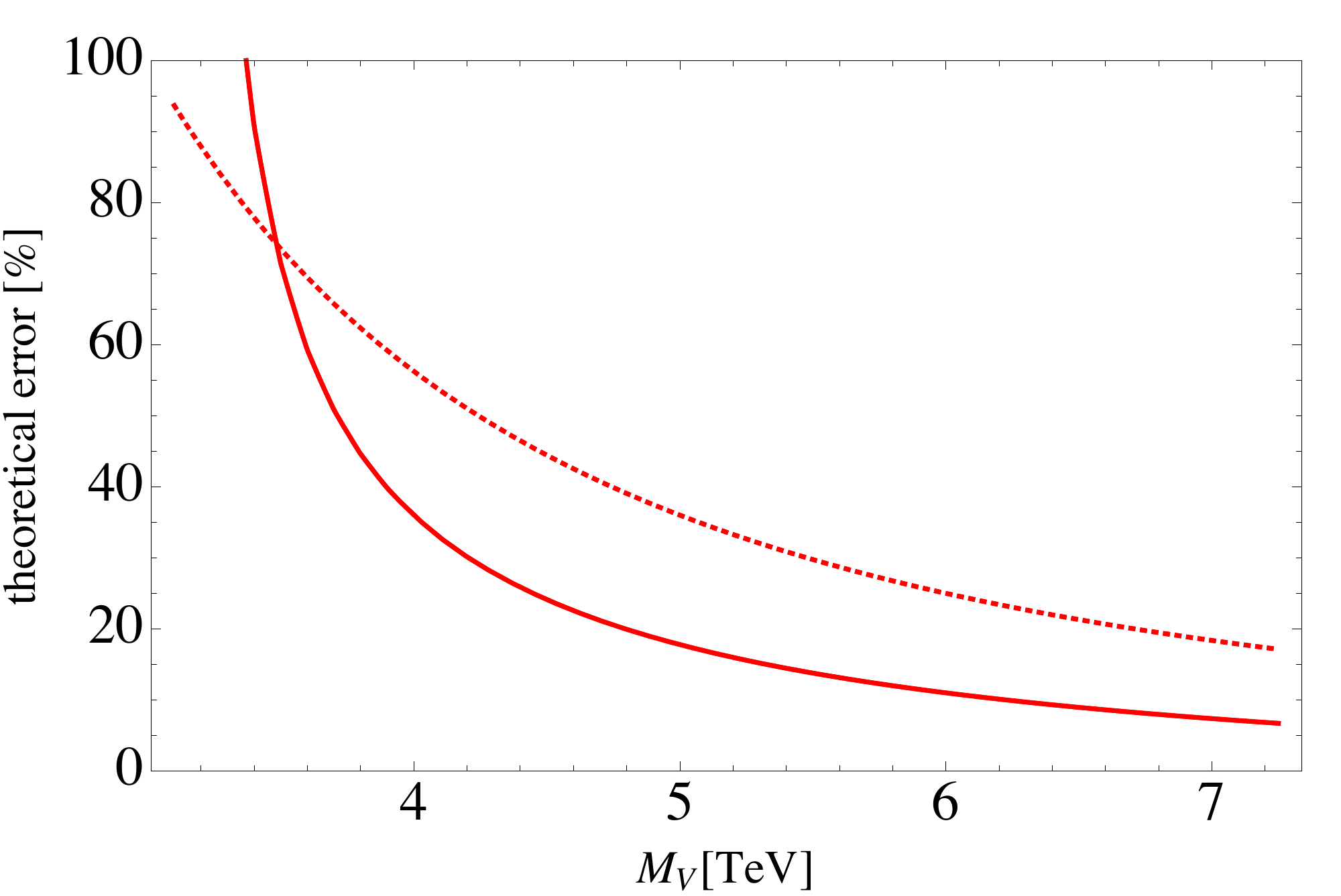} 
\caption{\emph{Left: The partonic $u \bar d \to W^+ h$ cross section as a function of  the center-of-mass energy of the parton collision. 
The black lines correspond to the $SU(2)_L$ triplet model with  $M_V =1\,$TeV,  $g_H=-g_q = 0.25$ (dashed),   $M_V = 2\,$TeV, 
$g_H=-g_q = 0.5$ (dotted), and  $M_V = 7\,$TeV, and $g_H=-g_q = 1.75$ (solid).  The corresponding  EFT predictions are shown 
in the linear approximation (solid red), and when quadratic terms in $D\!=\!6$ parameters are included in the calculation of the cross section (solid purple). 
Right:  Theory error as a function of $M_V$ (solid line).
The error is defined to be the relative difference between the constraints on $g_*^2 \equiv g_H^2 = g_q^2$ obtained by recasting the limits 
derived in the framework of a $D\!=\!6$ EFT and those derived from the resonance model.
The limits come from re-interpreting the hypothetical experimental  constraints with $M_{\rm cut} = 3\,$TeV, as  described in the text. 
The dotted line corresponds to the naive estimate $(M_{\rm cut}/M_V)^2$.
 }}
\label{fig:su2l}
\end{figure}

To illustrate this point, we compare the full and the effective descriptions of $q \bar q \to W^+ h$ for three benchmark points: 
\bi 
\item {\bf Strongly coupled:} $M_V = 7$~TeV, $g_H= -g_q = 1.75$; 
\item {\bf Moderately  coupled:} $M_V = 2$~TeV, $g_H=-g_q = 0.5$; 
\item {\bf Weakly  coupled:} $M_V = 1$~TeV, $g_H=-g_q = 0.25$\,.
\ei  
All three benchmarks lead to the same EFT parameters at the $D\!=\!6$ level.
However, because $M_V = \Lambda$ varies, these cases imply different validity ranges in the EFT. 
This is illustrated in Fig.~\ref{fig:su2l}, where we show (in the left panel) the production cross section as a function of
$M_{Wh}$, for both the full model and the EFT.
While, as expected, in all cases the EFT description is valid
near the production threshold, above a certain point $M^{\rm max}_{Wh}$ the EFT is no longer a good approximation of the UV theory. 
Clearly, the value of $M^{\rm max}_{Wh}$ is different in each case. 
For the moderately coupled case,  it coincides with the energy at which the linear and quadratic EFT approximations diverge. 
From the EFT perspective, this happens because $D\!=\!8$ operators can no longer be neglected. 
However, for the strongly coupled  case, the validity range extends beyond that  point. 
In this case, it is the quadratic approximation that provides a good effective description of the UV theory.  
As discussed in the previous section, that is because, for strongly-coupled UV completions, the quadratic contribution from $D\!=\!6$ operators dominates 
over that of $D \geq$ 8 operators in an energy range below the cutoff scale.

As an illustration of our discussion of setting limits on the EFT parameters and estimating the associated theoretical errors, consider the following example of an 
idealized measurement. Suppose an experiment  makes the following measurement of the $\sigma(u \bar d \to W^+ h)$  cross section at different values of $M_{Wh}$: 
{
\begin{center}
\begin{tabular}{c|c|c|c|c|c|c}
$M_{Wh}$[TeV] & 0.5 & 1 & 1.5  &  2 & 2.5 & 3 
 \\[0.1cm] \hline  &&&&&& \\[-0.45cm]
$\sigma/\sigma_{\rm SM}$ & $1 \pm 1.2$ & $1 \pm 1.0$  & $1 \pm 0.8$  & $1 \pm 1.2$ & $1 \pm 1.6$ & $1 \pm 3.0$  
\end{tabular}
\end{center}
}

This is meant to be a simple proxy for more realistic measurements at the LHC, for example measurements of a fiducial $\sigma(pp \to W^+ h)$ cross section in several 
bins of $M_{Wh}$.  For simplicity, we assume that the errors are Gaussian and uncorrelated.  These measurements can be recast as constraints on $D\!=\!6$ EFT 
parameters for different value of $M_{\rm cut}$, identified in this case with the maximum $M_{Wh}$ bin included in the analysis. For simplicity, in this discussion we only 
include $\delta g^{Wq}_L  \equiv [\delta g^{Zu}_L]_{11} - [\delta g^{Zd}_L]_{11}$
and ignore other EFT parameters (in general, a likelihood function in the multi-dimensional space of the EFT parameters should be quoted by experiments).  
Then the ``measured" cross section is related to the EFT parameters by
\beq
{\sigma \over \sigma_{SM}} \approx \left ( 1 + 160\, \delta g^{Wq}_L 
{M_{Wh}^2  \over {\rm TeV}^2}\right )^2\, .
\eeq 
Using this formula, one can recast the measurements of the  cross section as  confidence intervals on $\delta g^{Wq}_L$. 
Combining the $M_{Wh}$ bins up to $M_{\rm cut}$, one finds the following  $95\%$ confidence intervals:  
{
\begin{center}
\begin{tabular}{c|c|c|c|c|c|c}
$M_{\rm cut}$[TeV] & 0.5 & 1 & 1.5  &  2 & 2.5 & 3  
 \\[0.08cm]  \hline {\vrule height 16pt depth 10pt width 0pt}
$\delta g^{Wq}_L \times 10^{3}$ & [-70, 20] & [-16,4] & [-7,1.6] & [-4.1,1.1] &  [-2.7,0.8] & [-2.2,0.7] 
\end{tabular}
\end{center}
}
 \begin{figure}[ht]
\centering
\includegraphics[width=0.6 \textwidth]{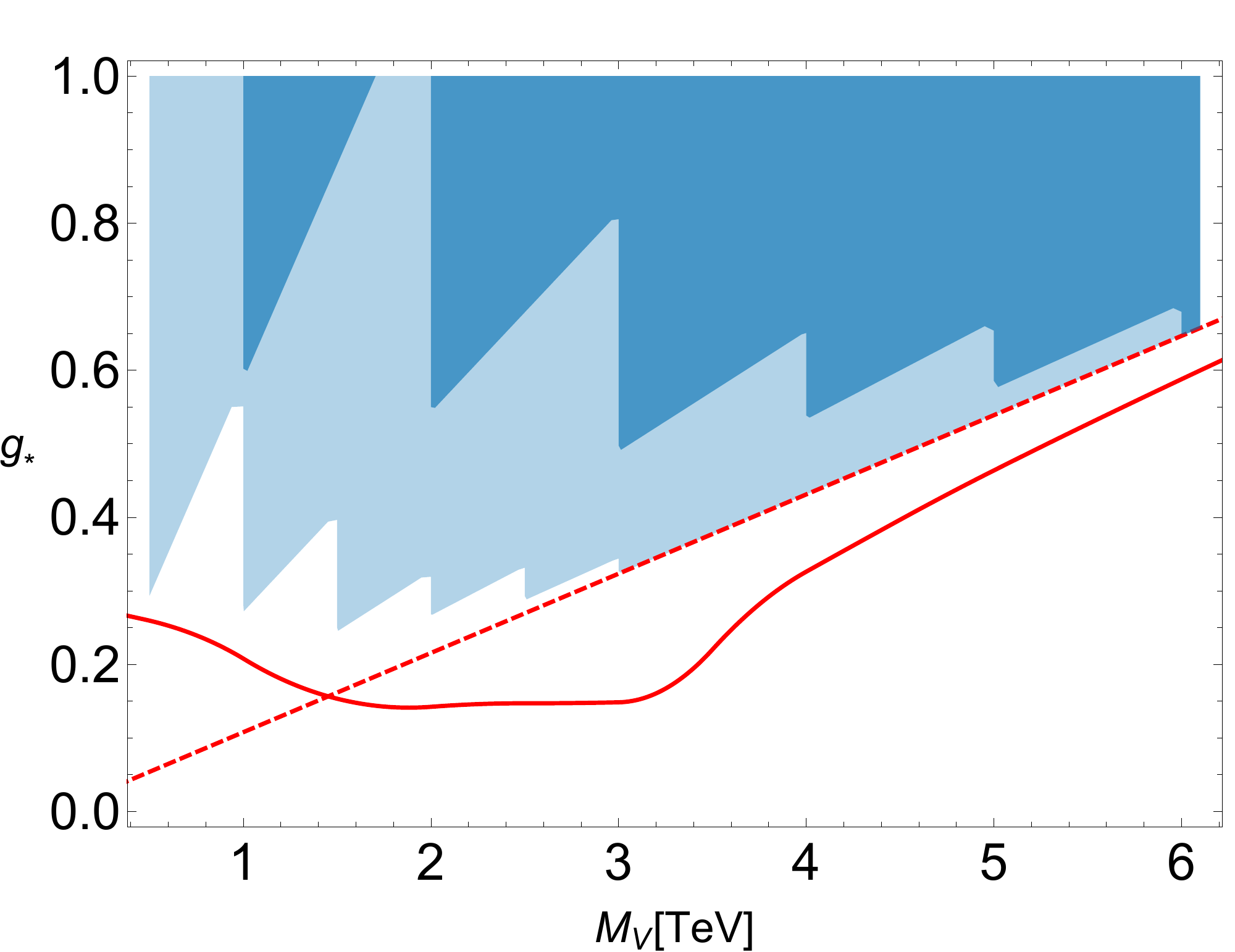}
\caption{\emph{{
Limits on the coupling strength $g_* \equiv - g_H = g_q$ as a function of the resonance mass~$M_V$. The solid and dashed red curves are obtained
respectively  from the vector resonance model and 
a naive analysis in terms of a $D$=6 truncated EFT, including data from all energies.
The dark (light) blue region corresponds instead to the bound derived from a consistent EFT analysis where only data with $M_{Wh}<M_{\rm cut}=\kappa M_V$ 
are used with $\kappa$=0.5 (1).
}
}}
\label{fig:gstar}
\end{figure} 

Suppose these constraints are the result of an experimental analysis.
A theorist may try to interpret them as constraints on the vector resonance model with $-g_q = g_H \equiv g_*$ using the map in Eq.~(\ref{eq:eftvsmodel}). 
The larger $M_{\rm cut}$ is, the stronger the limit  on $g_*$ will be for a fixed $M_V$.
For instance, by using the limits from the full dataset, 
$M_{\rm cut} = 3$~TeV,  one would obtain the constraint on $g_*$ 
given by the dashed red line in Fig.~\ref{fig:gstar}.
For large $M_V$ this approximates  well the  limits obtained by fitting the full BSM model to the same dataset (solid red line). 
In other words, for $M_V \gg 3$~TeV the theory error of the EFT is well under control, see the right panel in Fig.~\ref{fig:su2l}.
However, the difference between the EFT and the  true BSM limits increases as $M_V$ decreases. 
For $M_V \lesssim 3.5$~TeV, as the resonance enters the experimental reach,  the EFT limits have little to do with the true limits on the BSM theory; in other words the theory error explodes. 
However, it is still possible to obtain useful EFT limits in the low $M_V$ regime if the experimental results are quoted as a function on $M_{\rm cut}$.
In that case, for a given $M_V$, one can set a limit on $g_*$ using the data up to $M_{\rm cut} = \kappa M_V$, as in Eq.~(\ref{eq:bounds2}).  
The exclusion  obtained by such a procedure with $\kappa = 0.5 (1)$ is given by the dark (light) blue region in Fig.~\ref{fig:gstar}. 
Clearly,  for $M_V \gg 3$~TeV this procedure coincides with the usual EFT limit setting. 
On the other hand, for $M_V \lesssim 3.5$~TeV it returns a  consistent, though conservative limit on the resonance model. 
In other words, setting limits on $g_*$ at a given $M_V$ by using only the bins below $\kappa M_V$ allows one to keep the error
constant and of order $\sim \kappa^2$ even in the region at low $M_V$ where it would otherwise blow up.
Notice that the fact that the EFT constraint obtained with the full dataset (dashed line) matches the corresponding one in the exact theory (solid line) at 
$M_V\approx 1.5\,$TeV  is a mere accident of the  multi-bin analysis: for this value of $M_V$, the limit in the full model is dominated by the bin at $M_{Wh}=1.5\,$TeV, 
while that in the EFT is dominated by the highest bin at $M_{Wh}=3\,$TeV. 
As a matter of fact, the EFT limit is inconsistent for $M_V < 3\,$TeV and it gives an over optimistic exclusion in the region of very small resonance masses.

\section{Summary}

In this note we have discussed the validity of an EFT where the SM is extended by $D\!=\! 6$ operators. 
{We have emphasized} that the validity range cannot be determined using only low-energy information. 
The reason is that, while the EFT is valid  up to energies of order of the mass~$\Lambda$ of the new particles, 
low-energy observables  depend on the combinations $\tilde c^{(6)}/\Lambda^2$, where the
coefficient $\tilde c^{(6)}$ is a function of the couplings of the UV theory. 
We have pointed out that only when a particular power counting is adopted, for example the $g_*$-scaling discussed in this note, can the 
contributions from $D\!=\! 6$ and  $D\!=\! 8$ operators be estimated in a bottom-up 
approach,  and the error associated with the series truncation be established.
{In particular,  the power counting is  necessary to estimate the range of variation of the effective
coefficients and to identify when departures from the SM can be sizable, possibly bigger than the SM itself, compatibly with the EFT expansion.

The correction to a given observable from $D\!=\!8$ derivative operators  may be estimated to be of order
$(\tilde c^{(8)}/\tilde c^{(6)})(E/\Lambda)^2$ relative to that from $D\!=\!6$ operators, where $E$ is the relevant energy of the process.
The size of this effect depends on the dynamical features of the UV theory, {\it i.e.} on the value of $\tilde c^{(6)}$ and $\tilde c^{(8)}$.
In Section~\ref{sec:PC} we have discussed  the physical conditions determining the relative hierarchy among the effective coefficients.} 
{When the hierarchy between the $D\!=\!6$ and $D\!=\!8$ effective coefficients is entirely dictated by their scaling with $\Lambda$ (so that $\tilde c^{(8)}\approx\tilde c^{(6)}$),
the energy at which the EFT breaks down  coincides with the scale at which $D\!=\! 8$ and higher-dimensional operators become as important as $D\!=\! 6$ ones.
Conversely, when the EFT expansion is well convergent at the LHC energies,  the effects of $D\!=\! 8$ operators can be  neglected. Operators with more powers of the Higgs fields give relative corrections of order $(\tilde c^{(8)}/\tilde c^{(6)})(v/\Lambda)^2$, so that the validity of this second expansion does not depend on the energy of the process.}

Exceptions from this rule, in the form $\tilde c^{(8)}\gg\tilde c^{(6)}$, may arise in a controlled way as a consequence of symmetries and selection rules
or for certain well-defined classes of  processes, as we discussed in Section~\ref{sec:lim}.   
In these cases the $D\!=\! 8$ operators may give the leading correction at high energies and must be included to obtain a proper EFT description.
The inclusion of $D\!=\! 8$ operators in experimental analyses is justified only when dealing with these  special cases,  and would represent
an inefficient strategy in a generic situation.

Similar issues regarding the validity of the expansion may arise also in the context of LHC pseudo-observables. 
In Refs.~\cite{Gonzalez-Alonso:2014eva,Greljo:2015sla},  pseudo-observables are defined as form factors parametrizing amplitudes of physical processes 
subject to constraints from Lorentz invariance {(see also Ref.~\cite{David:2015waa})}.  
These form factors are expanded in powers of kinematical invariants of the process around the known poles 
of SM particles, assuming poles from BSM particles are absent in the relevant energy regime.  
{Such energy expansion is analog to the derivative expansion in the EFT, and a discussion parallel to the one presented in this note can be done about its validity. 
In particular, estimating the relative importance of the neglected higher-order terms always requires making assumptions on the UV theory, similarly to the EFT case. }


Besides discussing the range of validity of the EFT description, we have also identified situations where 1-loop insertions of $D\!=\!6$ operators
or higher-order real emissions can give a large effect and need to be included (see Section~(\ref{sec:loop})). At the practical level, if 
the size of some of the effective coefficients is constrained much less strongly (or measured to be much larger) than others, then it is possible 
that 1-loop effects mediated by these coefficients may be large.

If no large deviations from the SM are observed at the LHC Run-2, stronger constraints on $D\!=\!6$ operators can be set. 
As we discussed, this will extend the EFT validity range to a larger class of UV theories ({\it i.e.} those with smaller $c^{(6)}$) and
leave less room for contributions of $D\!=\!8$ operators.
As a consequence, the internal consistency and the validity range of the LO $D\!=\!6$ EFT will increase.\footnote{The validity range can also be improved by means of a global analysis combining different measurements,  which often lifts flat directions in the parameter space~\cite{Pomarol:2013zra,Falkowski:2014tna} and leads to stronger constraints on $D\!=\!6$ effective coefficients, {see {\it e.g.}~\cite{Falkowski:2015jaa,Butter:2016cvz}}.} 
In this regard, a different conclusion was reached by the authors of Refs.~\cite{Berthier:2015oma,David:2015waa}, although the discussion on the sources of theoretical 
errors presented in these papers agrees with ours.
We believe that the discrepancy is due to different assumptions on the underlying UV theory. In particular, Refs.~\cite{Berthier:2015oma,David:2015waa} discuss 
a situation in which both $\Lambda$ and $c^{(6)}$ are small, while $c^{(8)}$ is sizable. In our perspective this situation, rather than being generic,  
corresponds to one of the special cases discussed in Section~\ref{sec:lim}.

If, on the other hand, deviations from the SM are observed at the LHC Run-2, efforts to include EFT loop corrections and to estimate the effects of $D>6$ operators 
will be crucial to better characterize the underlying UV theory.
Notice that even in the case in which new particles are discovered in the next LHC runs, the EFT approach and the results 
presented here still remain useful. It may be indeed convenient to describe processes below the new physics threshold in terms of few effective operators
rather than the full set of new particles. Such low-energy studies can precisely extract properties of the SM fields and at the same time  
measure the coefficients of the effective  operators, generated by the new heavy resonances. Predictions in term of the full spectrum of new particles would instead rely on the knowledge
of their masses and couplings, which have to be extracted from high-energy data and might not be precisely determined.

Most of the discussion in this note is relevant at the level of the interpretation of the EFT results, rather than at the level of experimental measurements.  
However, there are also practical conclusions for experiments.  
We have proposed  a concrete strategy to extract bounds on (or determine) the effective coefficients of $D\!=\!6$ operators in a way which is automatically 
consistent  with the EFT expansion. This requires reporting the experimental results as functions of the upper cuts (here collectively denoted by $M_{\rm cut}$) 
on the kinematic variables, such as transverse momenta or invariant masses, that set the relevant energy scale of the process.
This is especially important for hadron collider experiments, such as those performed at the LHC, where collisions probe a wide range of energy scales.  
In general, knowledge of the experimental results as a function of $M_\text{cut}$ allows one to constrain a larger class of theories beyond the SM in a larger range 
of their parameter space.  An explicit example illustrating our procedure was given in Section~\ref{app:example}.
As a quicker (though less complete) way to get an indication on the validity range of the EFT description, it is also useful to present the experimental results both 
with and without  the contributions to the measured cross sections and decay widths that are quadratic in the effective coefficients.  
This gives an indication on whether the constraints only apply to strongly-interacting UV theories or they extend also to weakly-coupled ones.
Notice that even in situations where it makes sense to expand the cross section at linear order in the coefficients of $D\!=\!6$ operators, quadratic terms should 
always be retained in the calculation of the likelihood function, as we show in the Appendix. 
Other frameworks to present the results, as for example the template cross-sections, 
should also be pursued in parallel, as they may address some of the special situations discussed in this note.  
Finally, given its model-dependency, we suggest to report the estimated uncertainty on the results implied by the EFT truncation separately 
from the other errors, and to clearly state on which assumptions the estimate is based.

A concluding comment is in order when it comes to constrain explicit models from the bounds derived in an EFT analysis of the data.
Although EFT analyses aim at a global fit with all the operators included, it is important to ensure that the reported results are complete enough  
to later consider more specific scenarios  where one can focus on a smaller set of operators. Reporting the full likehood function, or at the very least the correlation 
matrix, would be a way to address this issue.

\section*{Acknowledgments}

We acknowledge comments and  feedback from the participants of the LHCHXSWG and in particular from
Christoph~Englert, Ayres~Freitas, Mart\'in~Gonz\'alez-Alonso, Admir Greljo, Fabio~Maltoni, Aneesh~Manohar, David~Marzocca, Jose-Miguel~No, Alex~Pomarol, 
Frank~Tackmann.  R.C. and F.R. would like to thank Riccardo~Rattazzi for useful discussions and suggestions.
The research of F.G. is supported by a Marie Curie Intra European Fellowship within the 7th European Community Framework Programme (grant no. PIEF-GA-2013-628224). C.G. is supported by the European Commission through the Marie Curie Career Integration Grant 631962 and by the Helmholtz Association.
The work of R.C. is partly supported by the ERC Advanced Grant No. 267985 Electroweak Symmetry Breaking, Flavour and Dark Matter: One Solution for Three Mysteries (DaMeSyFla).

\appendix
\renewcommand{\theequation}{A.\arabic{equation}} 
\setcounter{equation}{0}

\section*{Appendix}\label{app:likelihood}

In this appendix we discuss one technical issue regarding the contribution of $D\!=\!6$ and $D\!=\!8$ coefficients to the likelihood used to derive the results.
We have seen that when the UV theory is strongly coupled and the deviations from the SM are large, the $D\!=\!6$ squared terms dominates the cross section,
while $D\!=\!8$ ones are suppressed by a ratio of weak to strong couplings. The same holds true in computing the likelihood of course.
If instead the deviations from the SM predictions are small,~\footnote{This can occur either because the UV theory is weakly coupled or because, despite strong
coupling, the new physics scale is much higher than the energy probed by the experiment, {\it i.e.}, $\Lambda \gg 4\pi \sqrt{\hat{s}_{\rm max}}$.} 
the $D\!=\!6$ quadratic terms can be neglected in the cross section
but should be retained in the likelihood. This can be easily seen as follows.  
A cross section~$\sigma$ (or any other experimentally measured observable) can be schematically written as 
\begin{equation} \label{eq:xsec}
\sigma \simeq \sigma_{\rm SM} \left( 1 + 2 \frac{\delta^{(6)}}{A_{\rm SM}}  \tilde c^{(6)}  + 2 \frac{\delta^{(8)}}{A_{\rm SM}}  \tilde c^{(8)} + \left( \frac{\delta^{(6)}}{A_{\rm SM}}  
\tilde c^{(6)} \right)^2 +  \cdots  \right)
\end{equation}
where $A_{\rm SM}$ and $\sigma_{\rm SM}$ denote, respectively, the SM amplitude and SM cross section, while $ \delta^{(6)}  \sim O(E^2/\Lambda^2)$, and 
$\delta^{(8)} \sim  O(E^4/\Lambda^4)$ parametrize the effect of higher-dimensional operators. As before we make use of the dimensionless coefficients
$\tilde c^{(6)}_i = c^{(6)}_i \Lambda^2$, $\tilde c^{(8)}_i = c^{(8)}_i \Lambda^4$.
We have shown terms up to $O(1/\Lambda^4)$,  denoting those further suppressed with the dots.
The $\chi^2$ function (again, schematically) has the form: 
\begin{equation}\label{chi}
\begin{split}
\chi^2\propto (\sigma - \sigma_{\rm exp})^2 =  & \left ( \sigma_{\rm SM} - \sigma_{\rm exp} \right )^2 
+ 4 \sigma_{\rm SM}  \left ( \sigma_{\rm SM}- \sigma_{\rm exp} \right )   \frac{\delta^ {(6)}}{A_{\rm SM}} \tilde c^{(6)}   
  +  4 \sigma_{\rm SM}^2  \left(  \frac{\delta^{(6)}}{A_{\rm SM}} \tilde c^{(6)} \right )^2 \\[0.1cm]
 &  + 2 \sigma_{\rm SM}  \left ( \sigma_{\rm SM}- \sigma_{\rm exp} \right )   \left( 2 \frac{\delta^{(8)}}{A_{\rm SM}} \tilde c^{(8)}   
   +   \left(  \frac{\delta^{(6)}}{A_{\rm SM}}  \tilde c^{(6)}\right )^2 \right)
    +\cdots\,,
\end{split}
\end{equation}
where $\sigma_{\rm exp}$ is the experimentally measured value of the cross section, and the dots stand for $O(1/\Lambda^6)$ terms.
{Neglecting the last two terms in Eq.~(\ref{eq:xsec}) corresponds to dropping the second line of Eq.~(\ref{chi}).}
The dimension-8 term in the second line  enters formally at the same order $1/\Lambda^4$ as the one proportional to $(\tilde c^{(6)})^2$ in the first line, 
but it can be always neglected within the EFT validity regime where { $\tilde c^{(6)} \gg \tilde c^{(8)} E^2/\Lambda^2$.}
{Indeed, under the assumption of small deviations from the SM prediction, the multiplicative factor $(\sigma_{\rm SM}- \sigma_{\rm exp})$ is small and effectively 
scales like $1/\Lambda^2$.}
Similarly, the $(\tilde c^{(6)})^2$ term in the second line is multiplied by $(\sigma_{\rm SM}- \sigma_{\rm exp})$  and can be neglected in this regime.  
On the contrary, the $(\tilde c^{(6)})^2$ term in the first line  is not suppressed and in fact it should be retained to ensure that the $\chi^2$ has a local minimum.
It is also easy to show that including the term proportional to $\tilde c^{(8)}$ affects the best fit value of $\tilde c^{(6)}$ only by an amount of 
$O(E^2/\Lambda^2)$.~\footnote{Indeed, minimizing \eq{chi}  with respect to $\tilde c^{(6)}$, one finds, schematically,
\begin{equation}
\tilde c^{(6)}   \simeq  { \sigma_{\rm exp} - \sigma_{\rm SM} \over \sigma_{\rm SM} } \, \frac{A_{SM}}{\delta^ {(6)} } -  \tilde c^{(8)}  {\delta^ {(8)} \over \delta^ {(6)}}\, .
\end{equation} 
}
We thus conclude that while dimension-8 operators can be neglected, square terms from $D\!=\!6$ should be retained.

\bibliographystyle{JHEP} 
\bibliography{EFTvalidity_arxiv}

\end{document}